Iqra yousaf

Mphill Applied Chemistry

University of Engineering and Technology Lahore

Review Paper

# The Current and Future Perspectives of Zinc Oxide Nanoparticles in the Treatment of Diabtese Mellitus

Abstract: This review paper explores the synthesis, characterization, and biomedical applications of zinc oxide nanoparticles (ZnO NPs). The synthesis methods, including chemical and green synthesis, are thoroughly examined, highlighting their impact on nanoparticle properties. The ZnO NPs were characterized using XRD, FTIR, UV-Vis spectroscopy, and SEM, revealing their crystalline structure, functional groups, optical properties, and morphological features. The paper discusses the biological benefits of ZnO NPs, particularly their enhanced bioavailability and therapeutic effects due to their nanometric size. The antibacterial, anti-inflammatory, and antidiabetic activities of ZnO NPs are emphasized, noting their potential to improve glucose regulation and metabolic control. The synergistic effects of combining ZnO NPs with other therapeutic agents, such as drugs or plant extracts, are also explored, suggesting enhanced efficacy in treating diabetes and inflammation. Despite promising results, the paper acknowledges the need for more comprehensive studies to compare the effectiveness of ZnO NPs with other zinc compounds and to better understand their toxicity and therapeutic roles. The review concludes by calling for further research to optimize the use of ZnO NPs in biomedical applications and to fully exploit their potential benefits.

.Key words:
- ZnO: Zinc Oxide
- NPs: Nanoparticles
- XRD: X-Ray Diffraction
- FTIR: Fourier-Transform Infrared Spectroscopy
- UV-Vis: Ultraviolet-Visible Spectroscopy
- SEM: Scanning Electron Microscopy
- SGLT2: Sodium-Glucose Cotransporter
- T2D: Type 2 Diabetes
- MIL: Metformin and Insulin Combination Therapy
- FDA: Food and Drug Administration

1. **Introduction:**

Diabetes mellitus (DM) is a long-term disease where the blood sugar level remain consistently high [1]. Diabetes is generally categorized into three types: type 1 diabetes (T1DM), type 2 diabetes (T2DM), and gestational diabetes (GDM), with T2DM making up about 90% of cases globally. In T2DM, high blood sugar levels result from either insufficient insulin production by the pancreas or poor cellular response to insulin, known as insulin resistance. The development of T2DM is also influenced by the balance between the growth and death of pancreatic β cells [1, 2]. Diabetes mellitus not only lowers quality of life and life expectancy but also significantly increases the risk of serious complications. These can include issues like blindness, kidney failure, heart attacks, strokes, and the need for limb amputations. The global impact of these complications makes diabetes a critical healthcare challenge that demands urgent attention and solutions [3]. This condition is a metabolic issue resulting from either a lack of insulin, insulin resistance, or both [4]. Diabetes is a complex condition that needs a diverse treatment strategy. Because the disease is so intricate, anti-diabetic medications can be quite costly [5]. This is because current anti-diabetic medications are often used together with other drugs or insulin, which raises the total cost of treatment. The ideal goal is to develop a cost-effective anti-diabetic treatment that works through multiple mechanisms [3]. Type 2 diabetes mellitus (T2DM) is influenced by many factors, including multiple genes and environmental conditions. Research has identified more than 40 genes linked to a higher risk of T2DM. These genes play roles in various biological processes such as cell development and physiological functions. Environmental factors that contribute to the risk of T2DM include aging, obesity, and insufficient physical activity [6]. Chronic high blood sugar in diabetes can lead to serious complications like nerve damage, eye issues, kidney problems, and heart diseases. Additionally, T2DM often goes undiagnosed for a long time because many people do not show symptoms early on. As a result, about one-third to one-half of patients may only be diagnosed after developing complications related to their high blood sugar levels. Even though diabetes is very common, the precise processes that cause gestational diabetes, type 1 diabetes, and type 2 diabetes are still not completely understood [1, 2] . To create better ways to prevent, diagnose, and treat diabetes and its related complications, we need to gain a deeper understanding of what causes the disease.

**Causes of Disease Diabtese Mellitus**

1.1. Epigenetic in diabatese:
An epigenetic trait is a lasting characteristic that is passed down through generations, caused by changes in gene activity that don't affect the actual DNA sequence [7]. Key types of epigenetic changes include DNA methylation, modifications to histone proteins like methylation, phosphorylation, and acetylation, as well as regulatory RNAs [8]. Epigenetic modifications typically interact and regulate each other to form an epigenetic profile, which changes gene function by altering chromatin structure. This profile can either activate genes by loosening the chromatin to allow access to the DNA for transcription or silence genes by tightly packing the chromatin to prevent access to the DNA [9]. Epigenetic states are influenced by three types of signals: epigenators, epigenetic initiators, and epigenetic maintainers. The epigenator signal is triggered by environmental factors, initiating an intracellular response. This could involve modifications or protein interactions that activate the epigenetic initiator. For example, in plants, temperature can act as an epigenator. The intracellular epigenetic initiator then receives and transmits this signal, setting up the chromatin state at specific locations. Initiators can include non-coding RNAs, DNA-binding proteins, or any entities that recognize DNA sequences and establish the chromatin state. Epigenetic maintainers preserve this chromatin state across generations and throughout the cell cycle, but they cannot initiate it. An example of an epigenetic maintainer is the DNA methylation of CpG islands [7].

**DNA Methylation**
- Altered Gene Expressions:

In the mammalian genome, most CpG dinucleotides are methylated. The impact of DNA methylation on gene expression depends on its location within the gene's regulatory regions or gene bodies, with the effects varying based on the specific position of the methylation [11]. Methylation within the gene body is linked to reduced RNA polymerase transcription of the gene. Methylation in the promoter or first intron (regulatory region) is crucial for regulating transcription initiation and is generally associated with gene silencing. There are two non-exclusive models explaining this gene silencing: (1) methyl groups block the access of transcriptional co-activators to their target sequences, and (2) 5-methyl-CpG induces a repressive chromatin state [12]. Hypermethylation is often associated with gene silencing, while hypomethylation usually results in increased gene expression. Numerous studies have reported different levels of gene methylation in patients with type 1 diabetes [13-15] . The difference in methylation levels between type 1 diabetes (T1D) patients and healthy individuals has been identified at four CpG sites near the insulin gene, which encodes pre-proinsulin. This gene has the second highest odds ratio for increasing the risk of developing T1D. Specifically, hypomethylation at CpG-19, CpG-234, and CpG-135, as well as hypermethylation at CpG-180, have been linked to an increased risk of T1D [16]. The ratio of circulating methylated to unmethylated insulin genes has been linked to the risk of developing type 1 diabetes (T1D). Higher levels of methylated DNA from β-cells might serve as a potential biomarker for β-cell destruction in T1D [17]. A study involving 252 T1D patients and 286 healthy individuals found that T1D patients exhibited hypermethylation at CpG-373 and CpG-456 [18].

- **Environmental Factors**

Environmental factors such as diet, physical activity, and exposure to toxins play significant roles in altering DNA methylation, thereby influencing the risk of developing diabetes. Diet rich in nutrients like folate, B vitamins, and methionine can affect DNA methylation patterns. High-fat and high-sugar diets have been shown to lead to aberrant methylation of genes involved in glucose metabolism and insulin signaling, increasing diabetes risk. Conversely, diets rich in fruits, vegetables, and whole grains can promote beneficial methylation changes that protect against diabetes.

Physical activity also impacts DNA methylation. Regular exercise is associated with favorable methylation patterns in genes related to metabolic pathways, enhancing insulin sensitivity and glucose homeostasis. Physical inactivity, on the other hand, can lead to adverse methylation changes, contributing to metabolic dysfunction and increased diabetes risk.

Exposure to environmental toxins, such as heavy metals, air pollutants, and endocrine-disrupting chemicals, can disrupt normal DNA methylation processes. These toxins can induce hypermethylation or hypomethylation of critical genes, impairing their function and leading to insulin resistance and chronic inflammation, key factors in diabetes development.

- **Developmental Origins**

Epigenetic modifications during fetal development can significantly impact an individual's health, including their risk of developing diabetes later in life. These modifications involve changes to the DNA structure, such as the addition of chemical groups, that influence gene activity without altering the underlying genetic code.

Maternal nutrition plays a crucial role in shaping the fetal epigenome. For instance, if a pregnant woman has a diet deficient in essential nutrients, this can lead to changes in the fetal DNA methylation patterns. Such modifications can affect genes involved in metabolism and insulin production, increasing the child's susceptibility to diabetes.

Similarly, maternal stress during pregnancy can alter the fetal epigenetic landscape. Stress hormones, like cortisol, can cross the placenta and modify the expression of genes related to stress response and metabolic regulation in the fetus. These changes can predispose the child to metabolic disorders, including diabetes, by affecting how their body handles glucose and insulin.

In summary, the environment in the womb, influenced by maternal diet and stress, can lead to lasting epigenetic changes that affect gene function. These changes can predispose an individual to diabetes by altering key metabolic processes and insulin regulation, illustrating the profound impact of early developmental conditions on long-term health.

### 1.2 . Histone modifications

- **Chromatin structure**

Histone modifications, like acetylation and methylation, play a crucial role in chromatin structure and gene expression. Acetylation typically reduces the positive charge on histones, decreasing their affinity for DNA and resulting in a more relaxed chromatin structure. This relaxation allows transcription factors easier access to DNA, enhancing gene expression. Conversely, methylation can either activate or repress gene expression depending on the specific histone and amino acid involved. These modifications are vital in regulating genes related to insulin signaling and glucose uptake. For instance, increased acetylation of histones associated with the insulin receptor gene can enhance its expression, improving insulin sensitivity and glucose uptake. Conversely, repressive methylation marks can reduce the expression of genes critical for these pathways, potentially contributing to insulin resistance and metabolic disorders. Thus, histone modifications are key regulators of metabolic processes through their influence on chromatin dynamics and gene activity.

- **Inflamatory Pathways**

Epigenetic changes in histone modifications significantly impact genes associated with inflammation, playing a crucial role in the development of insulin resistance. Histones are proteins around which DNA wraps, and their modifications can either promote or inhibit gene expression. Specific modifications, such as acetylation and methylation, alter the accessibility of transcriptional machinery to DNA, thereby regulating gene activity.

In the context of inflammation, certain histone modifications can upregulate the expression of pro-inflammatory genes. For instance, increased histone acetylation at promoters of inflammatory cytokines can enhance their transcription. These cytokines, like TNF-α and IL-6, are known to interfere with insulin signaling pathways, leading to insulin resistance. Moreover, histone methylation at specific sites can either activate or repress inflammatory gene expression, further modulating the inflammatory response.

Chronic inflammation, driven by these epigenetic changes, can disrupt insulin receptor signaling through various mechanisms, including the impairment of insulin receptor substrate (IRS) proteins and the activation of stress kinases like JNK and IKK. This creates a feedback loop where inflammation exacerbates insulin resistance, contributing to metabolic disorders such as type 2 diabetes. Understanding these epigenetic mechanisms opens potential therapeutic avenues for targeting histone modifications to mitigate inflammation and improve insulin sensitivity.

### 1.3. Non-Coding RNAs

**MicroRNAs (miRNAs)**

MicroRNAs (miRNAs) are small non-coding RNAs that play a critical role in regulating gene expression by binding to messenger RNAs (mRNAs), leading to their degradation or inhibiting their translation. This regulation is crucial for maintaining normal cellular functions. In the context of diabetes, dysregulation of miRNAs has been implicated in insulin resistance and beta-cell dysfunction. Insulin resistance occurs when cells fail to respond effectively to insulin, a hormone essential for glucose uptake. Certain miRNAs can disrupt the signaling pathways that mediate insulin action, contributing to insulin resistance. Beta-cells, located in the pancreas, are responsible for producing insulin. When miRNAs are dysregulated, they can impair the function and survival of beta-cells, leading to inadequate insulin production. Thus, understanding miRNA dysregulation offers insights into the molecular mechanisms underlying diabetes and potential therapeutic targets for its treatment.

**Long Non-Coding RNAs(lnRNAs)**

Long non-coding RNAs (lncRNAs) are RNA molecules that don't code for proteins but play crucial roles in regulating gene expression. Unlike their protein-coding counterparts, lncRNAs function through various mechanisms to control when, where, and how genes are turned on or off. In the context of glucose metabolism and insulin signaling, lncRNAs have been shown to influence key genes involved in these processes. They can bind to DNA, RNA, or proteins, altering the structure of chromatin (the material that makes up chromosomes) and influencing the activity of genes. For example, some lncRNAs can enhance or suppress the transcription of genes by interacting with transcription factors or other regulatory proteins.

This regulatory capability is significant in glucose metabolism, where precise control of gene expression is essential for maintaining proper blood sugar levels. lncRNAs can affect how cells respond to insulin, a hormone crucial for regulating glucose uptake and utilization. Dysregulation of lncRNAs can lead to impaired glucose metabolism and contribute to conditions like diabetes.So,lncRNAs are essential players in the intricate network of gene regulation, particularly in processes critical for glucose metabolism and insulin signaling, highlighting their importance in health and disease.

**Transgenerational Effects**

Epigenetic modifications, such as DNA methylation and histone modification, can be inherited across generations, impacting the health of offspring. These changes do not alter the DNA sequence but affect gene expression, potentially influencing the risk of diseases like diabetes.One significant way these epigenetic marks are established is through environmental exposures and metabolic conditions experienced by parents. For instance, a parent's diet, stress levels, and exposure to toxins can lead to epigenetic changes that are passed to their children. During gametogenesis, when sperm and egg cells are formed, epigenetic modifications can be imprinted and subsequently transferred to the embryo. These changes can persist throughout the offspring's life and even into subsequent generations.

In the context of diabetes, if a parent experiences poor metabolic health or an unhealthy diet, this can lead to specific epigenetic changes. For example, high levels of glucose and fatty acids might alter methylation patterns on genes related to insulin production or glucose metabolism. When these altered patterns are inherited by the offspring, they may predispose them to insulin resistance or impaired glucose tolerance, increasing their risk of developing diabetes.

Animal studies have provided clear evidence of this phenomenon. For instance, mice exposed to a high-fat diet have shown epigenetic changes in their offspring, leading to increased susceptibility to metabolic disorders. Human studies, while more complex due to genetic and environmental interactions, also suggest that parental health can influence the epigenetic landscape of their children. For example, children born to mothers with gestational diabetes have a higher risk of developing type 2 diabetes later in life, potentially due to inherited epigenetic modifications.Thus, transgenerational epigenetic inheritance represents a crucial mechanism through which environmental and metabolic conditions can influence disease risk across generations, emphasizing the importance of parental health and environmental factors in the prevention of diseases like diabetes.

## 2. Importance of nanotechnology in medicine

Science has developed the ability to create materials at the nanoscale, known as nanoparticles. This nanoscale world has fascinated scientists, leading them to explore the structure and properties of objects at this tiny scale. Richard Feynman, who won the Nobel Prize in Physics in 1959, introduced the concept of "nano," and Norio Taniguchi was the first to use the term "nanotechnology" in 1974 [20]. Nanoparticles are tiny particles with sizes ranging from 1 to 1000 nanometers, though those between 1 and 100 nanometers are most commonly used in research. Nanoscience involves the study of physics, material science, and biology at the nanoscale, while nanotechnology focuses on observing, measuring, manipulating, and controlling materials at this tiny scale. Typically, nanoparticles have three layers: (a) the surface layer, which includes metal ions, polymers, and molecules; (b) the shell layer, which has a different chemical composition from the other layers; and (c) the core, which is the central part of the nanoparticle [31]. Nanobiotechnology is a rapidly growing field in clinical and medical research, gaining global popularity. This emerging field focuses on creating nanoparticles, which involves transforming bulk materials into much smaller particles. Nanomaterials are defined as materials, either manufactured or natural, that consist of unbound, aggregated, or clustered particles with external dimensions ranging from 1 nm to 100 nm [19] Due to their nanoscale size, nanoparticles exhibit unique behaviors and properties, such as antibacterial activity. Creating metal nanoparticles using physical and chemical methods has limitations due to their environmental impact and the requirement for costly equipment.[20] . Green approach technology is increasingly attractive because it is eco-friendly, non-toxic, and cost-effective. This approach focuses on using natural methods and is a major area of research, particularly in exploring various bioreducing agents. Plant phytochemicals and seed extracts play a unique role in nanoparticle synthesis, offering alternative therapeutic options in nanotechnology and nanomedicine.[21] . Herbs have been used as natural remedies since ancient times and are a key part of the traditional Indian system of medicine. Herbal plants are also regarded as a safer medicinal option compared to modern allopathic medicines [22]. The importance of using N. sativa (black cumin) is supported by scientific data on its safety, effectiveness, quality control, dosage, therapeutic uses, clinical trials, toxicity, and potential drug interactions. The practical use of N. sativa for medicinal purposes and commercial applications depends on the quality and yield of its seeds and essential oil. N. sativa is valued for its wide range of beneficial active phytochemicals. Combining nanoparticles with N. sativa could address various clinical issues. This review highlights the medicinal benefits of N. sativa seeds and explores their integration with different metallic or non-metallic nanoparticles for use in various medical and therapeutic applications [20]. N. sativa is a flowering plant from the Ranunculaceae family, native to northern India, Pakistan, and the Middle Eastern Mediterranean region. It is also commonly found in Arabia and Turkey. Known by names such as kala jeera, kalonji, and black cumin, it is also referred to as black caraway seeds, habbatu sawda, and habatul baraka ("Blessed Seed") [23]. The seeds also contain a significant amount of minerals such as zinc [20]. Studies have shown that treating diabetes-induced rats with N. sativa oil helps reduce oxidative stress, inflammation, and the amyloidogenic pathway. It also lowers the inhibitory effect of IOMe-AG538 on insulin receptors and alters the insulin-signaling pathway. As a result, it prevents neurotoxicity, the formation of amyloid plaques, and Tau hyperphosphorylation, while restoring normal levels of miRNA related to Alzheimer's disease [24]. In a meta-analysis of glycemic and serum lipid profiles, N. sativa supplementation was found to enhance fasting blood sugar levels, reduce glycosylated hemoglobin, and lower total cholesterol and LDL-cholesterol in patients with type 2 diabetes [25]. Diabetes results in high glucose levels, which causes the carbonyl group in glucose to react with amino acids in proteins, leading to harmful compounds known as advanced glycation end products. The effect of N. sativa extract on glycation was tested using glycated DNA samples analyzed through agarose gel electrophoresis. The results showed a significant reduction in glycation when N. sativa extract was present [26]. Thymoquinone has demonstrated potential as an agent that can help prevent glycation [27]. Some studies suggest that N. sativa can be used alongside oral antidiabetic medications as a supplementary treatment [28]. N. sativa plays an important role in managing inflammation and oxidative stress caused by diabetes. It offers several benefits, including restoring antioxidant defense systems, boosting antioxidant enzyme activity, reducing inflammatory biomarkers, suppressing pro-inflammatory mediators, and improving functions related to the endothelial system, liver, kidneys, heart, and immune system [29]. N. sativa and thymoquinone both help protect against endothelial dysfunction caused by diabetes. Tests with different doses of N. sativa have demonstrated its ability to inhibit the formation of advanced glycation end products (AGEs), even at lower concentrations. Thymoquinone was found to be the main compound responsible for this antiglycating effect [30].

## 3. Metallic Nanoparticles

Metal nanoparticles gained attention in the modern era when Faraday first identified them in a solution in 1857 and detailed their properties in his paper "Experimental Reactions of Gold (and Other Metals) to Light." However, the use of metals in medicine has a long history, dating back to the Vedic period in ancient India. The Ayurvedic scholar Acharya Charaka, from around 1500 B.C., described the medicinal properties of metals such as gold, silver, platinum, lead, and mercury in his work, the Charaka Samhita. Gold is the most commonly used noble metal for making nanoparticles, with silver being the second most popular. Other metals like iron, copper, tin, and lead are also used in nanoparticle synthesis. These metal nanoparticles exhibit various properties, including antibacterial, antioxidant, and optical effects. For instance, gold nanoparticles have demonstrated antioxidant properties [32] antibacterial, and antibiofilm [33] Gold nanoparticles are known for their antioxidant activities, whereas silver nanoparticles are recognized for their antibacterial properties [34].

## 4. Zinc Oxide Nanoparticles Importance in Diabetese Mellitus

Among the various types of nanoparticles, zinc oxide (ZnO) nanoparticles are significant industrial materials, and a substantial quantity of them is manufactured [35]. In recent years, ZnO nanoparticles have been created and tested as a new method for delivering zinc, exploring their potential antidiabetic effects in rats with diabetes induced by streptozotocin [36,37]. While these studies showed that ZnO has strong antidiabetic effects, the impact of this nanoparticle on oxidative stress parameters wasn't fully understood. Moreover, its therapeutic effectiveness compared to bulk forms hadn't been examined. Therefore, this study aimed to evaluate the potential adverse effects of ZnO in diabetic rats and compare the nanoparticle with zinc sulfate, focusing specifically on oxidative stress parameters [39-42]. The methods for producing stable ZnO nanoparticles have greatly improved over the years. The main techniques include solution-free mechano-chemical processes, chemical precipitation, sol-gel, solid-state pyrolysis, and biosynthesis [43,44]. Researchers are interested in metal-oxide nanoparticles due to their unique physical and chemical properties and their high surface area to volume ratio. These nanosized inorganic metal oxides demonstrate impressive biological activities even at very low concentrations and remain thermally stable across a wide range of temperatures [45-48]. Biosynthesized into various nanostructures such as nanotubes, nanoflowers, nanorods, and nanobelts, zinc oxide (ZnO) nanoparticles have garnered significant interest among researchers compared to other metal oxide nanoparticles. Zinc is an essential trace element involved in protein and nucleic acid synthesis, as well as other enzymatic processes. Consequently, the use of ZnO nanoparticles seems safe for healthy cells while being potent enough to induce apoptosis in cancerous cells [49-53]. Zinc oxide nanoparticles

(ZnO NPs) demonstrated antidiabetic potential by significantly increasing serum insulin levels and lowering blood glucose levels. Researchers also examined the gene expressions of insulin receptor A, GLUT-2, and glucokinase in diabetic livers, since zinc transporters, which play a key role in glucose metabolism, are present in both adipose tissues and the liver. In the liver, ZnO NPs significantly boosted the expression and activity of glucokinase [54]. ZnO nanoparticles (ZnO NPs) exhibit several effects that underscore their antidiabetic potential: (a) they increase insulin secretion and strengthen the antioxidant defense system in pancreatic β-cells; (b) they reduce blood glucose levels and improve glucose tolerance; (c) they enhance insulin signaling and sensitivity, as well as glucose uptake by the liver, skeletal muscle, and adipose tissue; (d) they inhibit lipolysis in fat cells and gluconeogenesis in liver cells. Additionally, due to their tiny size, orally ingested ZnO nanoparticles have a large surface area and more receptor sites, which enhances the drug's bioavailability. The combined action of the drug and the nanoparticles has a synergistic effect on diabetes treatment. The FDA has classified ZnO nanoparticles as a safe material [55]. Zinc oxide nanoparticles (ZnO NPs) were combined with drugs, and their synergistic effect was observed. Embaby et al. studied treatments using ZnO NPs or MIL, and the results showed improvement with only minor renal lesions and fibrosis. The combination therapies improved the parameters under study more effectively than either treatment alone [56] . There are still some drawbacks due to a lack of research in these areas: (1) comparing the biological benefits of zinc oxide nanoparticles with other metal nanoparticles; (2) the toxicity of zinc oxide nanoparticles towards biological systems remains a contentious issue; (3) evidence-based randomized studies specifically examining their therapeutic roles in enhancing anticancer, antibacterial, anti-inflammatory, and antidiabetic activities; and (4) understanding related animal studies regarding these properties [54]. Glifozins, also known as SGLT2 inhibitors, are a type of anti-diabetic medication that have a proven history of effectively treating type 2 diabetes (T2D) [57,58]. Glifozins are unique because they lower blood sugar levels in a way that's not dependent on insulin. They work by promoting the excretion of glucose through urine. This method of action not only helps control blood sugar but also improves various metabolic and blood flow issues that can increase the risk of cardiovascular disease. As a result, glifozins have different effects compared to other diabetes medications [59, 60].

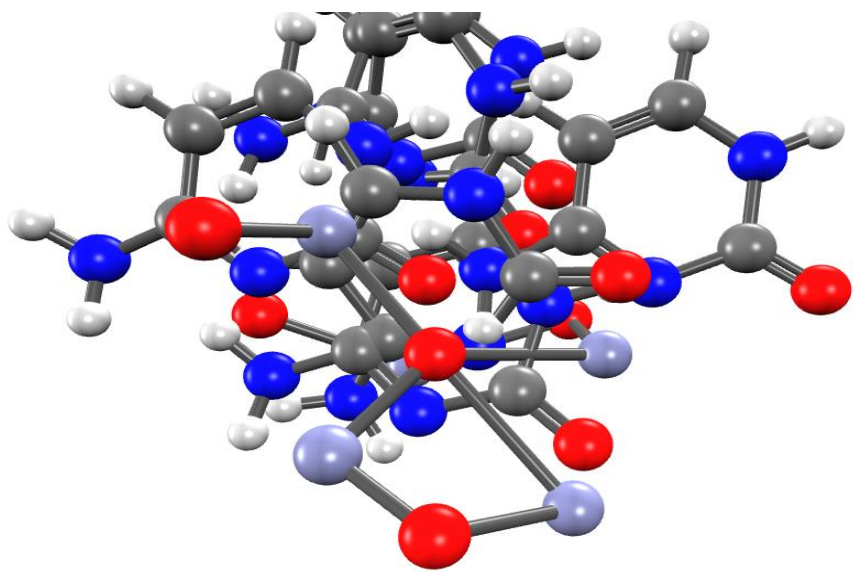

Molecular docking of zinc oxide nanoparticles with nitrogenous bases in nucleotides

Nitrogen atoms in nitrogenous bases

Oxygen atoms of zinc oxid molecules

Zinc atoms

## 5. Synthesis methods of zinc oxide

### 5.1. Chemical synthesis
we used only high-quality, analytical-grade chemicals without further purification. Zinc sulfate heptahydrate (ZnSO4·7H2O), which was used to make zinc oxide nanoparticles (ZnO NPs), and sodium bicarbonate (NaHCO3) were sourced from Merck. The main drug, empagliflozin, was obtained from Pharmagen Limited in Pakistan. Methanol, provided by Honeywell, served as the medium for drug adsorption, and deionized water was used for washing.

To prepare the zinc oxide nanoparticles (ZnO NPs), we first dissolved 1 gram of zinc sulfate heptahydrate (ZnSO4·7H2O) in 30 ml of bi-distilled de-ionized water and stirred the mixture with a magnetic stirrer until fully dissolved. We then added sodium carbonate solution (2 M) drop by drop until the pH reached 7, continuing to stir the mixture throughout.

After the pH was adjusted, we kept stirring the mixture for an additional hour. Next, we centrifuged and washed the product, then dried it at 120°C for about 12 hours. The dried material was ground into a fine powder that could pass through mesh screens. This powder was placed in a white crucible and calcined at 600°C for 2 hours, heating it at a rate of 10°C per minute [54].

### 5.2. Green synthesis of zinc oxide
The materials used included zinc sulfate heptahydrate (ZnSO4·7H2O) from Sigma-Aldrich, sodium hydroxide (NaOH) from Riedel-deHaën, distilled water as a solvent, and Nigella sativa seed extract. The Nigella sativa seeds were purchased from a local market. To prepare the seeds, they were cleaned with distilled water to remove any soil particles, then dried at room temperature, and ground into a fine powder.

Seventy grams of this powder were mixed with 250 ml of distilled water in a 500 ml beaker and boiled for 30 minutes. The mixture was then filtered through Whatman paper 01 to separate out the solid debris. The resulting 60 ml of extract was stored at 4°C for later use in synthesizing zinc oxide nanoparticles [59,60]. To prepare the zinc sulfate solution, zinc sulfate salt was dissolved in 40 ml of distilled water and mixed thoroughly using a magnetic stirrer. Various concentrations of zinc sulfate (0.18, 0.19, 0.24, 0.29, 0.35 M) were used to create different samples. Sixty ml of Nigella sativa seed extract was added to the solution drop by drop with a sterile syringe. The mixture was stirred continuously for at least 20 minutes. Next, a sodium hydroxide solution (0.50 M) in 50 ml of water was also added drop by drop, and the solution was stirred for 3 to 4 hours. The mixture was then centrifuged at approximately 3000 rpm for 15 minutes. The resulting zinc oxide nanoparticles were washed with distilled water to maintain a pH of 7 and then dried in a hot air oven at 150°C. After drying, the nanoparticles were ground into a fine powder using a mortar and pestle.

For antibacterial analysis, the laminar flow chamber and hands were sterilized with ethyl alcohol to avoid contamination. Petri dishes were sanitized in an autoclave at 121°C for 20 minutes. Luria Bertani Agar (LBA) medium was prepared by dissolving yeast extract, sodium chloride, agar, and tryptone in distilled water, which was then autoclaved at 121°C and 15 psi for 20 minutes. Once the LBA medium was set in Petri dishes, it was used for isolating and purifying bacterial strains.

## 6. Characterization Techniques

### 6.1. XRD and FTIR results
The crystal structure of the zinc oxide nanoparticles was analyzed using a D8 Discover X-ray Diffractometer with Cu Kα radiation. Fourier-transform infrared (FTIR) spectra were recorded with an IR Tracer-100 SHIMADZU spectrometer using the KBr pellet method. Optical absorption spectra were measured with a Hitachi U-2800 UV-Vis spectrophotometer across the 200–800 nm range. Dielectric properties and other parameters were determined using an Agilent E4980A LCR meter, which operates from 20 Hz to 2 MHz [62]. The structural analysis of the synthesized zinc oxide nanoparticles was performed by examining the X-ray diffraction (XRD) spectra, as shown in Figure 3. The XRD patterns revealed peaks at 31.66°, 34.42°, 36.24°, 47.43°, 56.69°, 62.78°, 67.96°, and 69.10° 2θ, corresponding to the (100), (002), (101), (102), (110), (103), (112), and (201) planes, respectively. These peaks match the hexagonal phase of ZnO nanoparticles as per JCPDS card no. 36-1451, with a space group of P63mc, and lattice parameters of a = 3.2498 Å and c = 5.2066 Å. The presence of sharp peaks indicates that the ZnO nanoparticles are highly crystalline.

The average crystallite size of the nanoparticles, determined using the Scherrer formula, ranged from 14 to 26 nm based on the XRD peaks. The Debye-Scherrer equation was used to estimate the crystallite size of the zinc oxide nanoparticles.

$$D_s = K\lambda / \beta \cos\theta$$

In the calculation of crystallite size ($D_s$), the constant $K$ is valued at 0.94, $\beta$ represents the full-width at half maximum (FWHM) of the diffraction peaks, and $\lambda$ is the wavelength of the Cu Kα radiation, which is 1.54 Å. The crystallite size refers to the dimensions of the individual crystals within a material and is typically measured using X-ray diffraction (XRD). This size plays a significant role in determining the material's properties, such as its mechanical strength, electrical conductivity, and optical characteristics [61]. As the concentration of ZnSO4 increases, the size of the crystallites also grows. This increase in size is influenced by the difference in ionic radii between $Zn^{2+}$ (0.074 nm) and $O^{2-}$ (0.140 nm), which contributes to the larger size of the nanoparticles [63] or by the high surface area [64]. The FTIR spectra of the ZnO nanoparticles, synthesized using Nigella sativa seed extract, were recorded in the range of 400 to 4000 cm−1 [65]. The FTIR bands observed between 399.30 and 514.99 cm−1 are attributed to the stretching vibrations of the Zn-O bonds in the ZnO nanoparticles [66]. The reduction of zinc ions to zinc oxide is indicated by specific bands in the spectra, which are due to the functional groups involved. These bands correspond to various stretching modes, each linked to a specific type of stretching. The presence of different bands suggests the formation of ZnO, with each band representing distinct stretching vibrations of the ZnO [67,68]. The broad bands observed between 500 and 800 cm−1 are attributed to vibrations of the Zn-O bonds. Metal oxides generally show absorption peaks in the range of 600 to 400 cm−1. The Zn-O frequencies found for the synthesized nanoparticles match those reported in the literature [69].

## 6.2. SEM Analysis

To examine the morphology of the synthesized ZnO at different concentrations, SEM analyses were conducted, as shown in Figure 6(a, b, c). These analyses were performed using a SEI-JEOL scanning electron microscope, with the system voltage set at 10 kV and a working distance of 10 mm between the piecing pole and the samples. The SEM analysis revealed pore morphology and irregular shapes, with some particles forming spherical clusters and others tending to aggregate. As the concentration of ZnSO4 increased, the particles showed more adhesion and a greater number of pores, as observed in Figure 6(c).

Grain size, which refers to the size of individual grains in a polycrystalline material, can significantly impact properties such as strength, ductility, fatigue resistance, corrosion resistance, and thermal conductivity. The grain sizes of the synthesized zinc oxide nanoparticles were calculated to be 0.45, 0.421, and 0.399 μm. It was observed that with increasing ZnSO4 concentration, the grain size decreased, leading to an increase in the surface area of the nanoparticles. The results show that the sizes and shapes of the nanoparticles are improved compared to existing literature on ZnO nanoparticles, likely due to differences in the synthesis method used [70].

## 6.3. Dielectric Study

The absorption of microwave energy by materials is influenced by the ability of the material's molecules to become electrically polarized. When a microwave electric field is applied, it can cause polarized molecules or atoms to rotate, transferring the microwave energy from the field into the material. The dielectric properties of materials are described by complex permittivity. The real part of complex permittivity, known as the dielectric constant, measures the energy stored in the dielectric material. The imaginary part measures the energy dissipated in the dielectric due to the applied electric field. To evaluate dielectric properties, sample pellets of the material are prepared with different concentrations and measured at room temperature. The dielectric constant of the material is determined based on the capacitance of the material, the thickness of the pellets, and the area of the pellets.The measurements indicate that the dielectric constant decreases with increasing frequency and is highest at lower frequencies. This reduction is due to the dipoles' inability to follow the variations in the electric field at higher frequencies. As the frequency increases, certain polarization mechanisms cease to function, causing an exponential decrease in the dielectric constant.The dielectric constant slightly increases with sample thickness and becomes constant at higher frequencies. This phenomenon is attributed to the relaxation effect in the space charge region, which arises due to the separation of high-conductivity phases and low-conductivity phases in the material's matrix. At lower frequencies, space charges can follow the applied electric field, but they fail to do so at higher frequencies [71-78].

## 7. Modulation of Glucose Metabolism

### 7.1. Moments Occur In Healthy Cells
1.Glucose-Stimulated Insulin Secretion (GSIS) Model (Panel A):
- Glucose Entry and Processing: Glucose enters the β cell through specific transporters (GLUT1/2/3) and is converted to glucose-6-phosphate (G6P) by glucokinase (GCK).  - ATP Production: G6P is then used in mitochondria to produce ATP.
- ATP-sensitive Potassium Channels (KATP): Increased ATP levels lead to the inactivation of KATP channels, which are made up of SUR1 and Kir6.2 subunits.
- Cell Membrane Changes and Calcium Influx: The closure of KATP channels causes the cell membrane to depolarize, activating voltage-gated calcium channels and allowing calcium to enter the cell.
- Insulin Release: The rise in intracellular calcium prompts the release of insulin from granules.
2. Genetic Influences on β Cell Function (Panel B):
- Key Proteins and Factors: The schematic shows various proteins and transcription factors crucial for β cell function, including those involved in cell cycle regulation (e.g., CDK4/6, cyclin D2) and ion channels (e.g., KvLQT1), as well as others like ZnT-8 that are important for zinc transport.
- Genetic Variants: It also highlights genes linked to genetic forms of diabetes and their impact on β cell function.

### 7.2. In Diabetic Cell
It seems that losing β cell mass or function leads to this unexpected change. Insulin might be needed to keep β cells' identity, and if less insulin is released in the local islet area, it could affect the identity of endocrine cells. For a long time, it has been known that too much α cell activity and the resulting high levels of glucagon contribute to high blood sugar in diabetic patients by increasing glucose production in the liver [79]. Therefore, tightly controlling the change from β cells to α cells (by reducing β cell flexibility) might help prevent diabetes from getting worse. In a mouse model, it was also shown that pancreatic α cells can change into β cells when PAX4 is overexpressed [80] or deletion of Arx [81,82]. In α cells, overexpressing PAX4 caused them to turn into β cells, leading to a decrease in α cells. This suggests new ways to restore β cell mass in diabetes. Not only can this provide new β cells, but it can also reduce the number of α cells, potentially restoring the balance between insulin and glucagon, which is disrupted in diabetes [83,84]. Besides α to β cell transdifferentiation, the pancreas can also regenerate by converting pancreatic ductal cells into β cells in rodents. Using a genetic lineage-tracing technique (a mouse model tracing duct-specific human carbonic anhydrase-II (CA-II)-positive cells with the CA-II promoter linked to the Cre-Loxp system) showed that CA-II-positive cells combined with β cells in the adult pancreas and ligated duct. This suggests that ductal cells might transform into new islet cells [85]. Additionally, studies have reported that rat and human pancreatic progenitor cells exist in the duct and have the potential to differentiate [86,87]. The ability of the endocrine pancreas to regenerate from ductal sources is further supported by our recent report, which shows increased growth in the pancreatic duct gland (PDG) area in humans with Type 1 Diabetes [88].

## 8. Preclinical Studies

**Animal Model and Induction of Diabetes**

Diabetic nephropathy (DN) is a leading cause of end-stage renal disease, characterized by progressive kidney damage resulting from diabetes mellitus. The search for effective treatments has led researchers to explore innovative therapeutic options, including the use of nanoparticles. Zinc oxide nanoparticles (ZnO-NPs) have emerged as promising agents due to their potential antidiabetic and protective properties.

In a recent study, researchers investigated the efficacy of ZnO-NPs in treating DN using a rat model induced by streptozotocin, a chemical that induces diabetes. The rats were divided into three groups: a control group, a DN group without treatment, and a DN group treated with ZnO-NPs. The ZnO-NPs were administered orally at a dose of 10 mg/kg/day for four weeks. The study's findings were promising. ZnO-NPs treatment led to significant improvements in both kidney function and structure. This was demonstrated by better preservation of the glomerular filtration barrier and nearly normal renal tissue structure. The beneficial effects of ZnO-NPs were attributed to their ability to maintain the function of podocytes, crucial cells in the kidney's filtration system, as evidenced by increased mRNA levels of nephrin and podocin. Additionally, ZnO-NPs appeared to activate autophagy, a cellular process that helps in the removal of damaged cells and proteins. This was indicated by enhanced expression of beclin-1 and LC3, proteins involved in autophagy, and reduced levels of p62, a marker of impaired autophagy. The activation of autophagy was associated with the inhibition of the mechanistic target of rapamycin (mTOR) signaling pathway, which is known to regulate cellular growth and metabolism.

Moreover, ZnO-NPs exhibited anti-apoptotic properties, reducing the expression of P53, a protein associated with cell death. The treatment also showed anti-inflammatory and antioxidant effects by decreasing the levels of cyclooxygenase-2 (COX-2), nuclear factor kappa beta (NF-κB), and hypoxia-inducible factors (HIF-1α), which are involved in inflammation and oxidative stress [89].

### 8.1. Experimental Groups and Protocol
**Group Allocation:**
Ten days after diabetes induction, rats were randomly assigned to three main groups in addition to a control group:
- Group Ic: Non-diabetic rats received a chow diet and 2 ml of 0.5% carboxymethyl cellulose sodium (a vehicle for Vildagliptin), with 10 rats in this group.
- Group IIc: Diabetic rats received a chow diet and 2 ml of 0.5% carboxymethyl cellulose sodium, also with 10 rats.
- Group III: Diabetic rats (n = 70) were further divided into 7 subgroups:
- Subgroup IIIa: Received Vildagliptin (10 mg/kg/day, oral) for seven weeks.
- Subgroups IIIb, IIIc, IIId: Received ZnONPs at doses of 1, 3, and 10 mg/kg/day, respectively.
- Subgroups IIIe, IIIf, IIIh: Received a combination of Vildagliptin (10 mg/kg/day, oral) and ZnONPs at doses of 1, 3, and 10 mg/kg/day, respectively.

Administration Method:
ZnONPs and Vildagliptin were administered via oral gavage for a duration of seven weeks.

### 8.2. Data Collection and Analysis
Body Weight, Water, and Feed Intake:
Measurements of body weight, water consumption, and feed intake were recorded weekly throughout the treatment period.
Sample Collection:
After the seven-week treatment period, rats were anesthetized intraperitoneally with Ketamine (50 mg/kg) and Xylazine (3–5 mg/kg). Blood was drawn from the abdominal vena cava, and the pancreas was excised and divided into two parts for biochemical and histopathological analyses.

### 8.3. Peripheral Blood Mononuclear Cell Isolation

**Blood Processing:**
Approximately 2.5 ml of blood from the rats were collected into EDTA tubes. Peripheral blood mononuclear cells (PBMCs) were isolated using a Ficoll density gradient (Ficoll-Hypaque, Sigma, St. Louis, MO), following the manufacturer's protocol.

### 8.4. Total RNA Extraction and Analysis
**RNA Extraction:**
Total RNA was extracted using TRIZOL reagent (Invitrogen, Carlsbad, CA), adhering to the manufacturer's instructions.

### 8.5. cDNA Synthesis and qPCR
The extracted RNA was converted to cDNA using Exiqon's miRCURY LNATM cDNA synthesis kit. MicroRNAs were poly-adenylated and reverse transcribed into cDNAs. Specific primers were used for amplification in a one-step SYBR Green I real-time PCR. The PCR conditions included an initial 10-minute incubation at 95°C, followed by 40 cycles of 95°C for 10 seconds and 60°C for 1 minute. The comparative ΔΔCt method was employed for data analysis, and unisp6 RNA spike-in was used for normalization.

### 8.6. Oral Glucose Tolerance Test

**Procedure**:
At the end of the experiment, rats were fasted overnight and then given 2 g/kg of a 50% glucose solution (0.5 g/ml) via gastric intubation. Blood glucose levels were measured at 0, 30, 60, 90, and 120 minutes post-glucose administration to assess glucose disposal and tolerance.

**Statistical Analysis**

Data were compiled and presented as means ± standard deviations. For the analysis of quantitative parametric variables, one-way analysis of variance (ANOVA) was employed, with subsequent post hoc comparisons performed using the Bonferroni test to evaluate differences between groups. Statistical analyses were conducted using SPSS software version 17 (Chicago, IL, USA). A significance level was set at $P < 0.05$[90].

The synthesis of nanoparticles using plant-based methods, commonly referred to as the "green method," is gaining recognition for its environmental and economic advantages over traditional chemical and physical methods. This approach utilizes plant extracts to facilitate the production of nanoparticles, such as zinc oxide nanoparticles (ZnO-NPs), in a manner that is both environmentally benign and cost-effective.

### 9. Environmental and Economic Benefits

The green synthesis method is considered environmentally friendly because it avoids the use of hazardous chemicals and toxic solvents typically involved in conventional nanoparticle synthesis processes. Instead, plant extracts, which contain natural compounds with reducing and stabilizing properties, are employed. These natural compounds facilitate the reduction of metal ions and the stabilization of nanoparticles, thereby minimizing the environmental footprint associated with the production of nanoparticles. Additionally, the use of plant extracts can significantly reduce the cost of synthesis. Plant materials are often abundant and inexpensive compared to the high-cost chemicals used in traditional methods. This makes the green synthesis approach not only more sustainable but also economically viable, particularly for large-scale production.

### 10. Antioxidant and Free Radical Scavenging Activities

ZnO-NPs synthesized through the green method using plant extracts have demonstrated noteworthy bio logical activities, including strong antioxidant and free radical scavenging properties. The antioxidant activity of these nanoparticles is attributed to the presence of bioactive compounds in the plant extracts used during synthesis. These compounds can include flavonoids, polyphenols, and other phytochemicals known for their ability to neutralize reactive oxygen species (ROS) and other free radicals.

The presence of these bioactive compounds in the plant extracts contributes to the enhanced antioxidant capacity of the resulting ZnO-NPs. This property is particularly valuable for various applications, including medical and cosmetic uses, where oxidative stress is a concern. For instance, in pharmaceutical applications, antioxidants can help mitigate cellular damage caused by oxidative stress, potentially offering therapeutic benefits in conditions related to oxidative damage.

In summary, the green synthesis of ZnO-NPs using plant extracts provides a more sustainable and economical alternative to traditional methods. The resultant nanoparticles not only align with environmental and cost-effective principles but also exhibit significant biological activities such as antioxidant and free radical scavenging, which broaden their potential applications in various fields [91, 92]. It is intriguing that zinc oxide nanoparticles (ZnO-NPs) synthesized via chemical methods did not demonstrate free radical scavenging activity. This lack of activity may be attributed to the differences in surface chemistry and morphology between chemically synthesized and green-synthesized nanoparticles. Chemical methods often produce nanoparticles with uniform sizes and shapes but may lack the bioactive surface features or stabilizing agents found in plant-extracted nanoparticles. These bioactive components, such as phytochemicals, in green-synthesized ZnO-NPs can contribute to enhanced antioxidant properties. Therefore, the chemical synthesis process might not impart the same level of functional groups or surface characteristics necessary for effective radical scavenging [91]. Zinc oxide nanoparticles (ZnO-NPs) synthesized using the green method demonstrated a notable reduction in cholesterol levels in alloxan-induced diabetic rats. This effect is likely due to the unique properties of ZnO-NPs produced by this environmentally friendly approach, which often involves the incorporation of bioactive compounds from plant extracts. These compounds may enhance the nanoparticles' ability to influence lipid metabolism and cholesterol levels. Additionally, the green synthesis process can result in nanoparticles with beneficial surface characteristics that interact more effectively with biological systems, potentially improving metabolic outcomes in diabetic conditions. Thus, the green-synthesized ZnO-NPs could offer therapeutic benefits by modulating cholesterol levels in diabetic rats [93]. The observed reduction in cholesterol levels in rats treated with green-synthesized ZnO-NPs was not mirrored in those administered chemically synthesized ZnO-NPs or insulin. This discrepancy suggests that the bioactive components present in the green-synthesized ZnO-NPs, which may enhance their interaction with metabolic pathways, play a crucial role in lowering cholesterol levels. However, both chemically synthesized and biologically prepared ZnO-NPs demonstrated superior efficacy compared to insulin in reducing fasting blood glucose levels. This indicates that ZnO-NPs, regardless of their synthesis method, possess significant antidiabetic properties, potentially offering advantages over insulin in managing blood glucose levels [93]. Further comparative research is necessary to explore the efficacy of ZnO-NPs synthesized from various medicinal plants. Such studies could reveal how different plant sources influence the biological activity and therapeutic potential of the nanoparticles. Additionally, recent investigations have focused on the toxicity and cellular uptake of ZnO-NPs compared to their bulk-size counterparts. These studies are crucial because the nanoscale form of ZnO-NPs can exhibit different biological interactions and toxicity profiles compared to larger particles. Understanding these differences is essential for evaluating the safety and effectiveness of ZnO-NPs in medical applications and ensuring their safe use in therapeutic contexts [94, 95]. Oral administration of zinc oxide nanoparticles (ZnO-NPs) and bulk ZnO can significantly influence the expression of genes associated with zinc transport in the mouse small intestine. This effect is likely due to the bioavailability of zinc from both forms of ZnO, which can interact with intestinal cells and trigger changes in gene expression. ZnO-NPs, with their increased surface area and reactivity, might have a more pronounced impact on gene transcription compared to bulk ZnO. These changes in gene expression could involve upregulation or downregulation of zinc transporters and other related proteins, reflecting the body's attempt to regulate and utilize the available zinc more effectively [96]. Zinc transporters, including zinc transporter 8 (ZnT8), are critical components in the β cells of the pancreas, where they play a vital role in insulin secretion. ZnT8 facilitates the transport of zinc into insulin granules, which is essential for the proper release of insulin. The importance of zinc in diabetes management has been highlighted by studies showing that zinc supplementation can benefit diabetic rats. Providing zinc to these animals helps to support insulin production and release, potentially improving glucose regulation and alleviating some symptoms of diabetes. This underscores the significance of zinc in maintaining pancreatic function and its potential therapeutic role in diabetes [97]. Zinc oxide nanoparticles (ZnO-NPs) have demonstrated superior antidiabetic effects compared to zinc sulfate, as evidenced by their ability to improve glucose clearance in STZ-induced diabetic rats. This enhanced antidiabetic activity is reflected in the observed increases in both serum zinc and insulin levels. ZnO-NPs likely offer a more effective form of zinc delivery due to their nanometric size, which may enhance bioavailability and cellular uptake. This leads to improved regulation of glucose levels and better overall metabolic control. In contrast, zinc sulfate may not be as efficiently absorbed or utilized, resulting in less pronounced effects on glucose metabolism and insulin secretion [98]. Zinc oxide nanoparticles (ZnO-NPs) synthesized using *Andrographis paniculata* leaf extract have demonstrated notable α-amylase inhibitory and anti-inflammatory activities. These properties are significant for managing diabetes and inflammation, as α-amylase inhibition can slow carbohydrate digestion, leading to reduced blood glucose levels, while anti-inflammatory effects can mitigate chronic inflammation associated with diabetes and other conditions. However, the efficacy of these green-synthesized ZnO-NPs in inhibiting α-amylase and exerting anti-inflammatory effects is somewhat less pronounced compared to zinc nitrate. Zinc nitrate, when used as a reference, tends to exhibit more potent activity in these areas due to its well-defined chemical properties and direct effects on biological systems. The slightly weaker effects observed with ZnO-NPs produced from plant extracts might be attributed to variations in the nanoparticle's surface chemistry, size, or the presence of additional stabilizing agents from the plant extract, which can influence their overall biological activity [99]. Despite the promising results of zinc oxide nanoparticles (ZnO-NPs) in managing diabetes, there is a notable scarcity of studies comparing their effectiveness with bulk ZnO and zinc salts in addressing various diabetic complications and symptoms. This gap highlights the need for more comprehensive research to fully understand how different forms of zinc-based therapies perform relative to each other in alleviating diabetic issues.

Synergistic effects between chemical compounds have long been a key focus in diabetes treatment. Combining different anti-diabetic drugs, which operate through distinct mechanisms, can enhance overall therapeutic outcomes. A notable example is the combination of sodium-glucose cotransporter 2 (SGLT2) inhibitors with metformin. This dual therapy has been shown to provide superior glycemic control compared to using either drug alone. The SGLT2 inhibitors reduce glucose reabsorption in the kidneys, while metformin enhances insulin sensitivity, leading to more effective management of blood glucose levels through their complementary actions. Such combination therapies underscore the potential benefits of exploring synergistic effects in the development of more effective treatments for diabetes [100]. In experimental diabetes models, synergistic interactions between zinc oxide nanoparticles (ZnO-NPs) and other therapeutic compounds, such as vildagliptin and thiamine, have been observed. These synergistic effects can enhance the overall therapeutic efficacy in managing diabetes. ZnO-NPs, known for their antioxidant and anti-inflammatory properties, can complement the actions of vildagliptin, a DPP-4 inhibitor that improves insulin secretion and reduces blood glucose levels. Similarly, thiamine, which supports glucose metabolism and reduces oxidative stress, may work effectively in combination with ZnO-NPs. Together, these combined treatments can provide a more comprehensive approach to controlling blood glucose levels and mitigating diabetic complications [101, 102]. The potential synergistic effects between zinc oxide nanoparticles (ZnO-NPs) and other antidiabetic medications or dietary supplements warrant further investigation. ZnO-NPs have demonstrated promising therapeutic benefits, including antioxidant and anti-inflammatory properties, which can complement the effects of conventional antidiabetic drugs. When combined with medications like vildagliptin or supplements such as thiamine, ZnO-NPs might enhance overall glucose control and mitigate diabetes-related complications more effectively than when used alone.

Exploring these synergistic interactions could reveal new avenues for optimizing diabetes treatment. Such research could help identify the most effective combinations and dosages, improving patient outcomes by leveraging the unique benefits of each component. Additionally, understanding these interactions could provide insights into how ZnO-NPs influence the pharmacodynamics of other antidiabetic agents, potentially leading to more tailored and effective treatment strategies. Thus, investigating these combined effects is crucial for advancing diabetes management and maximizing therapeutic efficacy.

**Conclusions**

Research into integrating zinc oxide nanoparticles (ZnO NPs) into nanofibers for diabetes treatment has yielded several key insights. ZnO NPs, due to their tiny size, significantly enhance the effectiveness and bioavailability of diabetes treatments. Incorporating these nanoparticles into nanofibers can lead to better glucose control and overall metabolic regulation, which are critical for effective diabetes management. Additionally, ZnO NPs are known for their antibacterial, anti-inflammatory, and antidiabetic properties. These characteristics can help reduce common diabetes complications, such as infections and chronic inflammation, thereby potentially improving patient health outcomes.

When ZnO NPs are combined with other therapeutic substances, like drugs or plant extracts, the results can be even more effective. This synergistic approach can boost the overall efficacy of diabetes treatments, offering a more comprehensive solution for managing the disease. While the initial findings are promising, more extensive studies are needed to compare the effectiveness of ZnO NPs with other zinc compounds. Additionally, it is crucial to understand the long-term safety and therapeutic potential of ZnO NPs to ensure their safe use in medical treatments.

Future studies should focus on optimizing the application of ZnO NPs in medicine. This includes exploring different methods of synthesizing these nanoparticles to enhance their properties and investigating how they interact with biological systems to maximize their benefits. In summary, incorporating ZnO NPs into nanofibers is a promising strategy for improving diabetes treatment, offering enhanced therapeutic effects and significant health benefits. However, further research is essential to fully understand their safety and effectiveness, paving the way for their optimized use in medical applications.